ARTICLE   OPEN

# Strong phonon localization in PbTe with dislocations and large deviation to Matthiessen's rule

Yandong Sun[1], Yanguang Zhou[2], Jian Han[3], Wei Liu[1], Cewen Nan[3], Yuanhua Lin[3], Ming Hu[4]* and Ben Xu[1,3]*

Dislocations can greatly enhance the figure of merit of thermoelectric materials by prominently reducing thermal conductivity. However, the evolution of phonon modes with different energies when they propagate through a single dislocation is unknown. Here we perform non-equilibrium molecular dynamics simulation to study phonon transport in PbTe crystal with dislocations by excluding boundary scattering and strain coupling effect. The frequency-dependent heat flux, phonon mode analysis, and frequency-dependent phonon mean free paths (MFPs) are presented. The thermal conductivity of PbTe with dislocation density on the order of $10^{15}$ m$^{-2}$ is decreased by 62%. We provide solid evidence of strong localization of phonon modes in dislocation sample. Moreover, by comparing the frequency-dependent phonon MFPs between atomistic modeling and traditional theory, it is found that the conventional theories are inadequate to describe the phonon behavior throughout the full phonon spectrum, and large deviation to the well-known semi-classical Matthiessen's rule is observed. These results provide insightful guidance for the development of PbTe based thermoelectrics and shed light on new routes for enhancing the performance of existing thermoelectrics by incorporating dislocations.

## INTRODUCTION

Thermoelectric devices directly convert waste heat into reusable electricity[1] and have a promising future, even though their efficiency is low. An effective approach to enhance the power factor is to design materials that consist of defects such as substitutional impurities, vacancies, and micro-nanostructural precipitations, which hinder the propagation of phonons while preserving the electronic transport ability.[2,3] Recently, the introduction of dislocations into thermoelectric materials to reduce the thermal conductivity (κ) to an extremely low level is found to be suitable to increase the efficiency.[4–7] As phonons carry most of the heat in thermoelectric materials, phonon-dislocation scattering is apparent; however, it is still an open question how the phonons of different energies interact with the local distortions induced by dislocation. Pioneers of these studies have tried to understand this effect from phenomenological ways. For example, Callaway assumed that the phonon scattering process can be investigated by frequency-dependent relaxation times and proposed a model based on the Debye approximation to calculate the κ.[8] Furthermore, the scattering effect was separated into two parts, namely, the distributions from the dislocation core (DC) and dislocation strain (DS), where the phonon relaxation time followed: $\tau_{DC}^{-1} \propto \omega^3$ and $\tau_{DS}^{-1} \propto \omega$.[9] Even though this theory is in reasonably good agreement with the experimental results by the fitting parameters,[4–7] the microscopic picture of the phonon-dislocation interaction is still missing. Ninomiya tried to use a second quantization to interpret the interaction of the dislocation strain field with the thermal activated phonons,[10] which provided a basic physical picture of the phonon-dislocation scattering. This was validated by the work of Li et al.,[11] however, only the dislocation strain field from the quantum theory was considered, which excluded the contribution from the dislocation core. Wang et al. carried out an ab initio approach based on Green's functions to include this contribution.[12] However, due to the scale of DFT, only a very small dislocation dipole model was considered where the contribution from an isolated dislocation was hardly described and was far from the real dislocation. Thus, it is both interesting and important to understand the interaction between phonons and a single dislocation from the point of view of basic physics as well as the applications.

In this study, non-equilibrium molecular dynamics (NEMD) simulations were carried out to study phonon transportation across an edge dislocation in a rock salt PbTe crystal. A comparison was made between the dislocation model (D Model) and perfect model (P Model), the only difference between the two models being the absence of an extra atom plane of edge dislocation in the P Model. First, the temperature distribution along the heat flux was directly obtained from the NEMD simulations. The frequency-dependent contribution of the heat flux in both the models was then determined by the recently proposed frequency domain direct decomposed method (FDDDM).[13–15] From that, the MFPs of the phonons were fitted according to the length-dependent spectral heat flux by performing the NEMD simulations with models of different lengths. In addition, the eigenmodes of a dislocation dipole model were analyzed to demonstrate the phonon behavior induced by the dislocation and finally the results were compared with the theoretical predictions of the traditional Debye–Callaway model.

[1]Laboratory of Advanced Materials, School of Materials Science and Engineering, Tsinghua University, Beijing 100084, People's Republic of China. [2]Mechanical and Aerospace Engineering Department, University of California Los Angeles, Los Angeles, CA 90095, USA. [3]State Key Laboratory of New Ceramics and Fine Processing, School of Materials Science and Engineering, Tsinghua University, Beijing 100084, People's Republic of China. [4]Department of Mechanical Engineering, University of South Carolina, Columbia, SC 29208, USA. *email: hu@sc.edu; xuben@mail.tsinghua.edu.cn

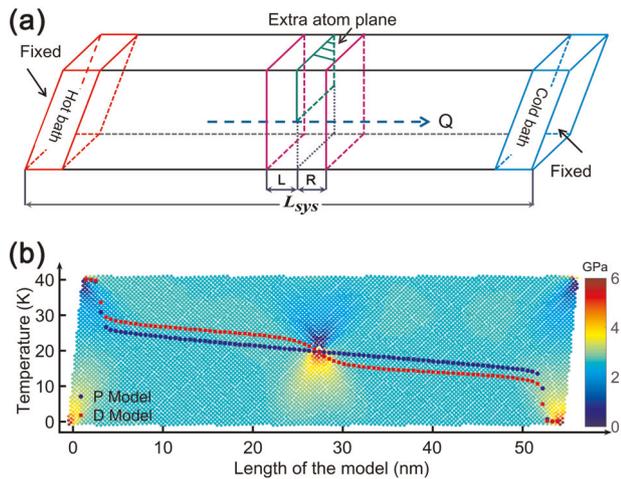

**Fig. 1** **a** Schematic of simulation cell used in NEMD: atoms at both ends of the system are fixed. Next to the fixed boundary, atoms within the length $L_{bath} = 2$ nm in the left and right side are coupled to Langevin heat baths, 40 K in the hot bath and 0.1 K in the cold bath, respectively. Two groups of atoms (denoted by L and R) are selected on both side of the extra atom plane (green color) from the edge dislocation and their velocities are sampled at successive time steps which are used to calculate the spectral heat flux. The same simulation procedure is done for the P Model. **b** Corresponding temperature profile of P Model and D Model and the normal stress field $\sigma_{xx}$ of D Model, colored by the $\sigma_{xx}$

**Table 1.** The thermal conductivity $\kappa_L$ calculated by Fourier's law from NEMD simulations

|  | Heat flux $J_Q$ (GW/m$^2$) | Temperature gradient $\partial T/\partial x$ (K/nm) | Thermal Conductivity $\kappa_L$ (W/(m·K)) |
| --- | --- | --- | --- |
| D Model | 4.350 | 0.4885 | 8.904 |
| P Model | 5.237 | 0.2248 | 23.30 |
| D/P Ratio | 83.06% | – | 38.21% |

## RESULT AND DISCUSSION

The simulation geometry is shown in Fig. 1a. A temperature gradient is applied on the longest dimension of the model. The details of the model parameters and NEMD simulation are given in Supplementary Note 1 (See Supplementary information at [URL will be inserted by publisher] for [Dislocation model details, FDDDM details, phonon eigenmode annlysis, calculation of participation ratio (PR) and theoretical phonon MFPs of Debye–Callaway model]) and in the method part. The temperature distribution is shown on the strain field of the dislocation sample in Fig. 1b. A linear temperature gradient is obtained for the P Model; however, a temperature drop is found around the dislocation area in the D Model. The temperature drop is about 6.5 K (16.25% of the total temperature difference). The temperature gradient is nonlinear near the reservoirs, as the maximum phonon MFP is longer than the length of the models, and hence, the phonons are scattered between the hot and cold reservoirs.[16] Two important interferences must be eliminated in the study of the phonon-dislocation scattering. The first is the strain field interaction between two dislocations which is overcome by the large distance between two dislocations and the second is the scattering at the free surface or boundaries, which is eliminated by introducing periodic boundary conditions. In this case, the temperature distribution is shown for an isolated dislocation and the temperature drop is mainly caused by this dislocation. By applying Fourier s law, the $\kappa_L$ of D Model is reduced significantly by 61.79% due to the dislocations. The result is shown in Tables 1, 23.30 W/(m·K) for P Model and 8.904 W/(mK) for D Model at 20 K. Moreover, we carried out separate MD simulations at room temperature and obtained thermal conductivity of 2.85 W/mK for bulk PbTe using the same classical potential, which is quite close to the experimental value.[17] The separate study proves that the interatomic potential used for our MD simulation is accurate enough and also all our MD simulation results are correct.

Therefore, it is important to identify the cause of the drastic temperature drop and the part of the phonon transportation that is prohibited by the dislocation. Moreover, there are no related results available in previous researches to the best of the authors' knowledge. In this study, the frequency-dependent heat flux $q(\omega)$ is investigated by the FDDDM method. The spectral heat flux through the cross section containing the half atom plane of the dislocation is shown in Fig. 2a. The following are the observations: (i) the overall tendency corresponds to the phonon density of states (DOS) (a perfect model with $6 \times 6 \times 6$ unit cell using method in Supplementary Note 3 (See Supplementary information at [URL will be inserted by publisher] for [Dislocation model details, FDDDM details, phonon eigenmode annlysis, calculation of PR and theoretical phonon MFPs of Debye–Callaway model])), especially in the low frequency mode where the phonons have a longer wavelength than the scale of the model and are not scattered. (ii) The number of phonons of particular frequencies is considerably decreased by the dislocation. $q(\omega)$ is decreased in the regions from 1.3 to 2.4 THz and 3.0 to 3.9 THz in the D Model. The difference in the heat flux clearly shows the reduction. An abrupt reduction is observed in the frequency range from 2.1 to 2.4 THz, which is due to the deviation of the frequency of the phonons caused by the stress of the dislocation. Previous DFT calculations have shown that in-plane tensile stress can decrease the acoustic phonon frequency.[18,19] (iii) The $q(\omega)$ at frequencies lower than 1.3 THz changes slightly between the P and D Models. This can be understood by considering that low frequency phonons have relatively long wavelengths and are usually behave as ballistic transport rather than diffusion in our finite length models. Thus, these low frequency phonons will not be scattered by the dislocation and pass through with no reduction in energy. Once the frequency-dependent heat flux $q(\omega)$ are determined, the accumulated $\kappa_L$ can be determined by dividing it by the temperature gradient and making an integration. The results is 9.0 W/(m K) for D Model and 19.6 W/(m K) for P Model, shown in Fig. 2b, which is close to our NEMD results.

The following part stresses on the atomic analysis of phonon behavior in the dislocation core region which leads to the significant decrease in the value of $\kappa_L$. Phonon eigenmode analysis is carried out for the dislocation dipole model as well as the corresponding perfect model in the PbTe crystal by using lattice dynamics with pair-wise potential (see Supplementary Note 3 (See Supplementary information at [URL will be inserted by publisher] for [Dislocation model details, FDDDM details, phonon eigenmode annlysis, calculation of PR and theoretical phonon MFPs of Debye–Callaway model])). Several representative eigenmodes at Γ point are shown in Fig. 3a–f (drew in OVITO[20]). Figure 3a, d shows the eigenmodes of the low frequency phonons and the atom vibrates in several groups, which is even disorderly, in the dislocation dipole model instead of a collective movement throughout the perfect model. For the medium frequency phonons, the eigenmodes are shown in Fig. 3b, e, where the atoms appear to exhibit random vibrations in the dislocation dipole model. Figure 3c, f demonstrate the localized vibrations of the dislocation core whose frequency is higher than the cut-off phonon frequency of the perfect crystal. These reveal that the high frequency phonons in the dislocation model tend to localized. Therefore, the PR of the eigenmodes,[21] (see

Y. Sun et al.

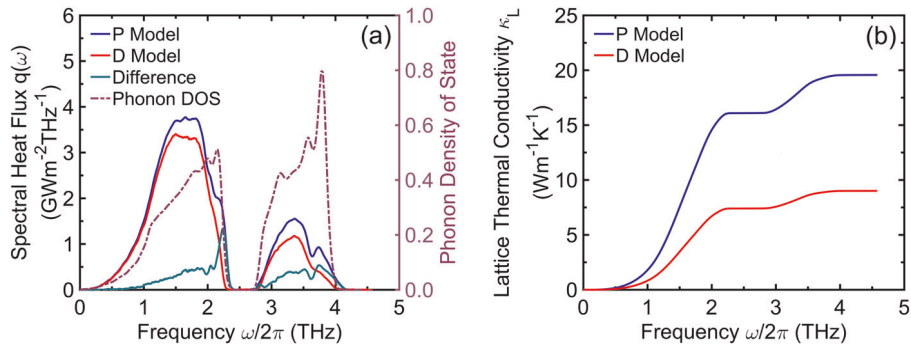

**Fig. 2** **a** Frequency-dependent spectral heat flux of P Model, D Model, and their differences, denoted by blue, red, cyan solid lines. The purple dash line is the phonon density state (DOS) of a perfect crystal. **b** The accumulated thermal conductivity of P Model and D Model, denoted by blue and red lines

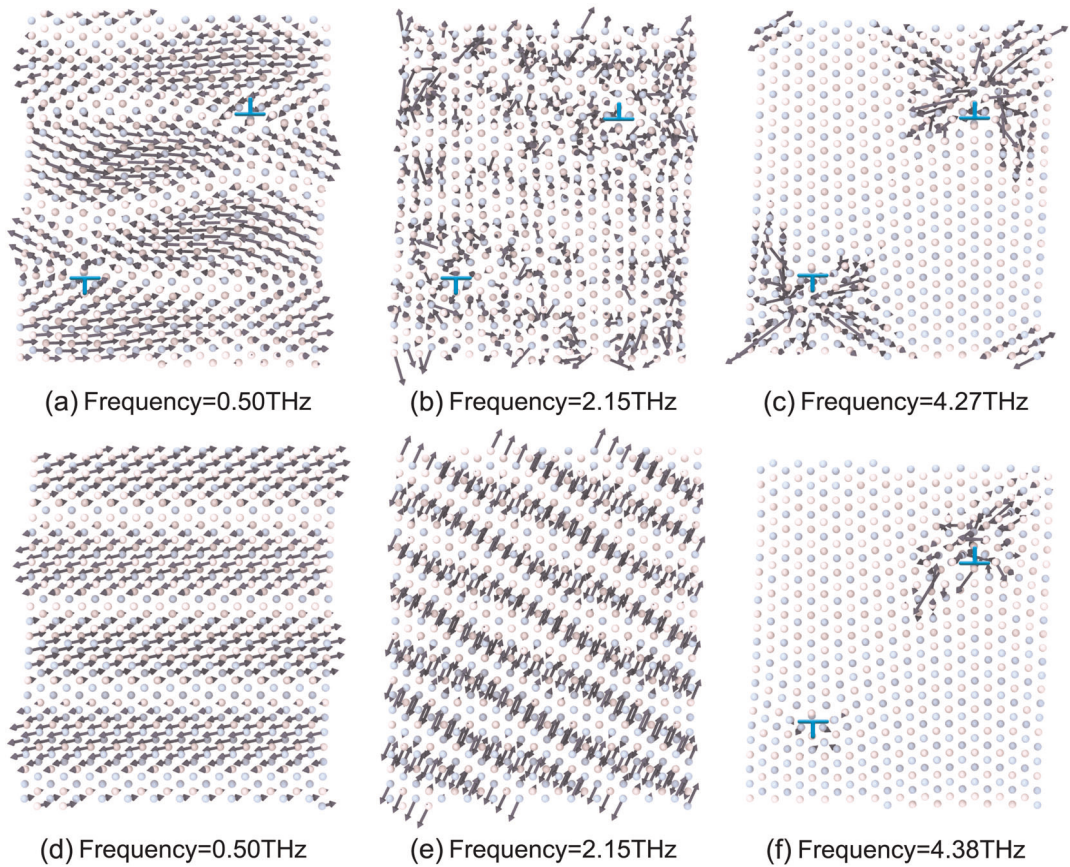

**Fig. 3** Phonon modes of PbTe models at different frequencies, magnitude and direction are shown by arrows: (**a**–**c**, **f**) for D Model, (**d**, **e**) for P Model. Pb and Te atoms are indicated by light cyan and pink color respectively

Supplementary Note 4 (See Supplementary information at [URL will be inserted by publisher] for [Dislocation model details, FDDDM details, phonon eigenmode annlysis, calculation of PR and theoretical phonon MFPs of Debye–Callaway model]), which signifies the degree of atoms participating in a given mode and is independent of the system size, is determined. Its value changes from $O(1)$ for spatially extended modes to $O(1/N)$ for a mode completely localized in a single atom.

The values of the PRs of the dislocated models reduce drastically as compared to the corresponding perfect models, as shown in Fig. 4a. The eigenvalues of the non-trivial portion of these modes decrease to below 0.1, most of which are at high frequencies and generated by the vibrations of the dislocation core. Vibration modes with such low values of PR were observed earlier and termed locons.[22] The existence of locons broadens the phonon spectrum which increases the possibility of phonon scattering. In addition, the value of the PR in the dislocation model is significantly reduced for modes of all frequencies as compared to that in the perfect model. Localized vibrations, where the atoms vibrate only in local coordination, are unable to deliver thermal energy as efficiently as the delocalized ones, and therefore decrease the $\kappa$ in the dislocation model.[23]

As a result of the strong localization of the phonon modes, the phonon transportation in the dislocation model is greatly

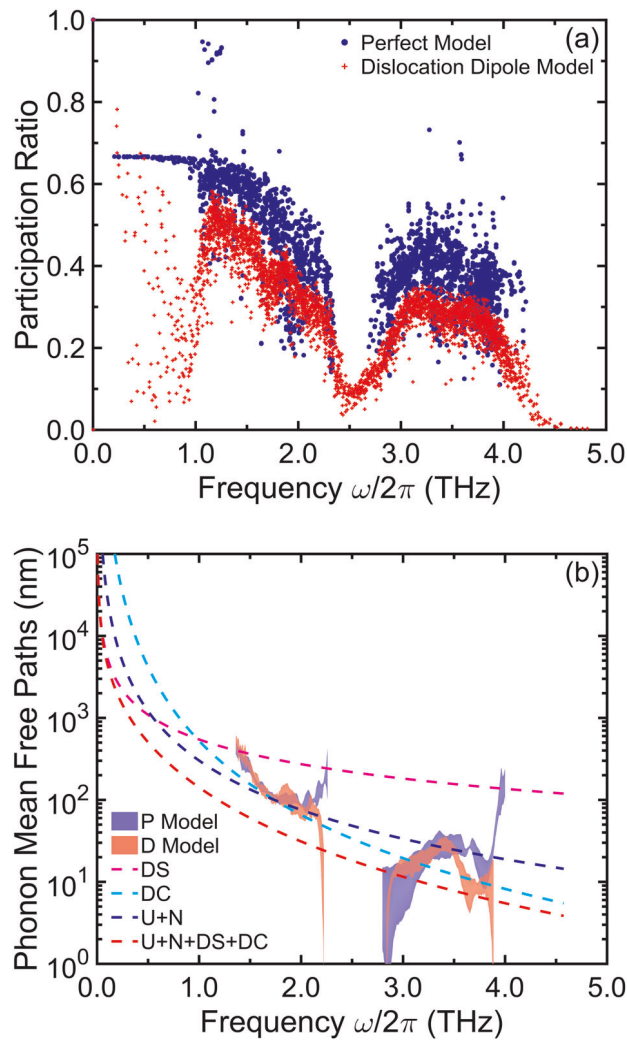

**Fig. 4** **a** Comparison of participation ratio between dislocation dipole model and its corresponding perfect model in PbTe crystal. **b** Phonon MFPs with 95% confidence interval from fitting the length-dependent $q(\omega, L)$ to Eq. (1) and theoretical prediction of phonon MFPs based on Debye–Callaway model with different contributions

scattered, which can be demonstrated by the MFP analysis of the dislocation sample. Based on the $q(\omega, L)$ from FDDDM, the phonon MFPs $\Lambda(\omega)$ are fitted according to the length-dependent $q(\omega, L)$ from the Lorentz equation[24,25] which was successfully used to describe the length-dependent heat flux in various systems.[24–27]

$$q_{(\omega,L)} = \frac{q^0(\omega)}{1 + L/2\Lambda(\omega)}, \quad (1)$$

where $q^0(\omega)$ denotes the number of modes at frequency $\omega$, corresponding to the ballistic transmission. The frequency-dependent spectral heat flux in models of different lengths $q_{(\omega,L)}$ is plotted in Supplementary Fig. 1 (See Supplementary information at [URL will be inserted by publisher] for [Dislocation model details, FDDDM details, phonon eigenmode annlysis, calculation of PR and theoretical phonon MFPs of Debye–Callaway model]). The system length $L_{sys}$ is changed by enlarging the fixed regions in the same model. As expected, increasing the system length reduces the heat flux due to the increased phonon scattering. However, ballistic transport can be clearly observed when $f \leq 1.3$ THz which suggests that the phonon MFPs in this frequency range are longer than the system dimensions under consideration. The ballistic spectral heat flux and phonon MFPs with 95% confidence intervals are obtained by fitting $q^0(\omega)$ and $\Lambda(\omega)$ to Eq. (1). The part of the curves where $q_{(\omega,L)}$ is almost the same or nearly equal to zero are neglected to avoid large errors. Therefore, the MFPs are not distributed over the full range of frequency.

The realistic MFPs in the dislocation sample are compared to those obtained from phenomenological models which are based on the Debye–Callaway model and include the Umklapp process scattering, normal process scattering, dislocation strain, and dislocation core scattering, shown in Fig. 4b. The fitting details are given in Supplementary Note 5 (See Supplementary information at [URL will be inserted by publisher] for [Dislocation model details, FDDDM details, phonon eigenmode annlysis, calculation of PR and theoretical phonon MFPs of Debye–Callaway model]). The difference between the blue and red dashed lines shows the proportion of phonon-dislocation scattering. The dislocation strain scattering (DS) dominates the low frequency part whereas the dislocation core scattering (DC) dominates the high frequency part, which agrees with the results obtained in previous studies. The realistic phonon MFPs decrease with increasing phonon frequency in all the models. Figure 4b shows that phonons with frequencies above 2.8 THz have short MFPs below 100 nm whereas phonons with frequencies between 1.2 and 2.3 THz have longer MFPs ranging from 100 to 1000 nm. The phonon MFPs in the frequency ranges from 2.0 to 2.2 THz and 3.4 to 3.9 THz are significantly reduced due to the localization of the phonon modes, as mentioned above, hence, logarithmic coordinate axes are used. Although the theoretically predicted MFPs of the phonons from Matthiessen's Rule agree reasonably well with the simulation results, there is still a large deviation. According to theoretical predictions, phonons in the entire frequencies range are scattered to a certain degree by the dislocation, however, only phonons of a certain frequency range are scattered by the dislocation in the calculations. It is expected that the dislocation will scatter phonons of different frequencies in different materials.

In conclusion, boundary scattering and strain coupling effects were excluded in the study of the phonon-dislocation scattering. The influence of a single edge dislocation on thermal transport in a PbTe crystal was investigated and it was shown that the $\kappa$ calculated by the Fourier s law in the NEMD simulation decreases by 62% with a $4 \times 10^{15}$ m$^{-2}$ dislocation density. The result upholds previous experimental studies[4,5] which show that a $10^{15}$ to $10^{16}$ m$^{-2}$ dislocation density is necessary to reduce the $\kappa$ by introducing dislocations. By applying the FDDDM method, the frequency range of the phonons that are most scattered by dislocations was determined. Through the eigenmode analysis, it was found that the phonons close to the dislocation become localized and, therefore, decrease the $\kappa$. By comparing the calculated and theoretically predicted MFPs of the phonons, it was established that the theoretical predictions were inadequate as the scattering rate of phonons of different frequencies from the dislocation was only expressed as a simple function of frequency. Thus, the proposed method has the advantage of a quantitative scaling of the scattering rate of phonons from the dislocation and sheds light on the application of heat management of materials.

## METHODS

### NEMD simulation

The NEMD simulations are carried out in LAMMPS package[28] with a timestep of 5 fs. A Buckingham potential taken from ref. [29] is used to model the interactions between atoms. We perform $4.5 \times 10^6$ NEMD steps, corresponding to a total running time of 22.5 ns. The first 7.5 ns is used to relax the structure with isothermal-isobaric ensemble (NPT) and canonical ensemble (NVT), then the following 10 ns is used to obtain a steady temperature gradient and heat flux. The hot region and cold region are coupled to Langevin heat baths with damping parameter 0.1 (20 fold of the timestep of the simulation) and keyword tally yes. The region between them is coupled to microcanonical ensemble (NVE). The velocity field information for FDDDM calculation and the average temperature are sampled in the last 5 ns.

### Frequency domain direct decomposed method (FDDDM)
To quantitatively clarify phonon-dislocation scattering processes, spectral heat flux $q(\omega)$ through the dislocation is investigated by our in-house FDDDM code, see the Supplementary Note 2 for full details.

### Phonon eigenmode analysis
The phonon DOS and eigenmodes are calculated by Phonopy software[30] combined with classical molecular dynamics force constant, see the Supplementary Note 3 for full details.

### DATA AVAILABILITY
The datasets generated during and/or analyzed during this study are available from the corresponding author on reasonable request.

### CODE AVAILABILITY
The codes used in this study are available from the corresponding author on reasonable request.

### ACKNOWLEDGEMENTS
The authors thank Prof. G. Jeffery Snyder of Northwestern University for helpful discussions and Dr. TianShu for polishing the language. Simulations were performed with computing resources granted by the National Supercomputer Center in Tianjin under project TianHe-1(A). Research reported in this publication was supported by the Basic Science Center Project of NSFC under grant No. 51788104 and by the Tsinghua National Laboratory for Information Science and Technology. Research reported in this publication was supported in part by the NSF and SC EPSCoR/IDeA Program under award number (NSF Award # OIA-1655740 via SC EPSCoR/IDeA 19-SA06). The views, perspective, and content do not necessarily represent the official views of the SC EPSCoR/IDeA Program nor those of the NSF.

### AUTHOR CONTRIBUTIONS
Y.D.S. performed the MD simulations, data processing, and manuscript writing. Y.G.Z. developed the FDDDM code. J.H., W.L., C.W.N., Y.H.L., M.H. and B.X. participated in discussion and interpretation of results and contributed to outlining the review paper.

### COMPETING INTERESTS
The authors declare no competing interests.

### ADDITIONAL INFORMATION
**Supplementary information** is available for this paper at https://doi.org/10.1038/s41524-019-0232-x.

**Correspondence** and requests for materials should be addressed to M.H. or B.X.

**Reprints and permission information** is available at http://www.nature.com/reprints

**Publisher's note** Springer Nature remains neutral with regard to jurisdictional claims in published maps and institutional affiliations.